\documentclass[12pt,eqsecnum,nofootinbib,floats,article,aps,prd,floatfix,titlepage,tightenlines]{revtex4} 
\usepackage{epsfig}

\usepackage{graphics}
\usepackage{bm}

\begin{document}
\title{Bounces with O(3)$\bm \times$O(2) symmetry}

\author{Ali Masoumi}
\email{ali@phys.columbia.edu}
\author{Erick J. Weinberg}
\email{ejw@phys.columbia.edu}
\affiliation{Physics Department, Columbia University, New York, New York 10027
\vskip 0.5in}

\begin{abstract}

We study the contribution to the decay of de Sitter vacua from bounces
with O(3)$\times$O(2) symmetry.  These correspond to the thermal
production of a vacuum bubble at the center of a horizon volume with
radius $r_H$ and a temperature defined by the horizon.  They are
analogues of the flat spacetime bounces, independent of Euclidean
time, that correspond to thermal production of a critical bubble.  If
either the strength of gravity or the false vacuum energy are
increased, with all other parameters held fixed, the bounces approach,
and eventually merge with, the Hawking-Moss solution.  Increasing the
height of the barrier separating the true and false vacuum, and thus
the tension in the bubble wall, causes the center of the bubble wall
to approach, but never reach, the horizon.  This is in contrast with
the prediction of the thin-wall approximation, which inevitably breaks
down when the wall is near the horizon.  Our numerical results show
that the Euclidean action of our solutions is always greater than that
of the corresponding O(4)-symmetric Coleman-De Luccia bounce.

\end{abstract}

\maketitle

\section{Introduction}

Vacuum decay by quantum tunneling in curved spacetime was first
studied more than thirty years ago by Coleman and De Luccia
(CDL)~\cite{Coleman:1980aw}, using a formalism based on bounce
solutions of the Euclidean field equations.  In recent years there has
been renewed interest in this topic, motivated in part by its relevance
for a possible string landscape.  This has uncovered new aspects of
the problem that become significant if the mass scales approach the
Planck mass or the bounce radius approaches the horizon length.

It was assumed in~\cite{Coleman:1980aw} that the dominant bounce
solution had an O(4) symmetry; we will refer to such solutions as CDL
bounces.  In this paper we will investigate a different class of
solutions, with an ${\rm O(3)}\times{\rm O(2)}$ symmetry, that can
occur when the tunneling is from a de Sitter spacetime.

Our motivation for studying these is two-fold.  For the case of
tunneling in flat spacetime at zero temperature it has been shown that
the bounce with minimum action in a theory with a single scalar field
has an O(4) symmetry, and it is generally believed that this result
extends to theories involving more fields.  Indeed, it is quite
possible that all bounce solutions in these theories are
O(4)-symmetric.  No proof of a comparable result in curved spacetime
has ever been obtained.  The ${\rm O(3)}\times{\rm O(2)}$ symmetry
that we impose (which would not be possible in flat spacetime at zero
temperature) reduces the field equations for the bounce to a set of
ordinary differential equations, thus making the problem quite
tractable and allowing comparison of the actions of the two types of 
bounces.

A second motivation is that these solutions shed light on a
reinterpretation of the CDL prescription that was advocated recently
by Brown and one of us~\cite{Brown:2007sd}.  It was argued there that
in the case of decay from a de Sitter vacuum the bounce should be
understood as describing tunneling in a finite horizon volume at
finite temperature.  This has implications for extracting from the
bounce the configuration immediately after tunneling.  The CDL bounce
is topologically a four-sphere, and it is commonly asserted that the
configuration after tunneling is obtained from the three-sphere lying
on the ``equator'' of the bounce.  By contrast, it was argued in
Ref.~\cite{Brown:2007sd} that half of this three-sphere gives the
configuration within the horizon volume after tunneling, with the
other half giving the configuration before tunneling.  The fields
beyond the horizon, either before or after tunneling, are never
specified and do not enter the calculation.  For CDL bounces
corresponding to vacuum bubbles much less than horizon size there is
little practical difference between the two interpretations.  The
differences become significant as the bubble approaches horizon size
and, especially, in the Hawking-Moss~\cite{Hawking:1981fz} limit.  For
the bounces we consider, the differences in interpretation are
manifest for all bubble sizes.  From the viewpoint of
Ref.~\cite{Brown:2007sd} these bounces correspond to thermal
production of a single bubble, whereas the conventional viewpoint
would yield the somewhat surprising conclusion that a pair of bubbles
had been produced.

Our interest in this these bounces was first raised by the work of
Garriga and Megevand~\cite{Garriga:2003gv,Garriga:2004nm}.  Motivated
by the analogy with the thermal production of bubbles in flat
spacetime, they studied bounces with ${\rm O(3)}\times{\rm O(2)}$
symmetry and, indeed, did interpret these as corresponding to the
nucleation of a pair of bubbles.  The discussion was originally in the
context of brane nucleation~\cite{Brown:1987dd,Brown:1988kg}, in which
the bubble walls are closed 2-branes, but their analysis can also be
applied to field theory nucleation in the thin-wall approximation.  We
will differ from them in not restricting ourselves to this thin-wall
limit.  In fact, we will see that the thin-wall approximation
inevitably breaks down when the bubble becomes sufficiently large, in
what is perhaps the most interesting limit of these solutions.

We work in the context of a scalar field theory with Lagrangian
\begin{equation}
    {\cal L} = \frac12 (\partial_\mu \phi)^2 - V(\phi)  \, ,
\end{equation}
where $V(\phi)$ has two local minima, with the higher (lower) being
termed the false (true) vacuum.  We assume that both minima have
positive energy, and are thus de Sitter spacetimes.  Although our
focus is on the nucleation of a true vacuum bubble in a false vacuum
background, we also consider the nucleation of a bubble of false
vacuum  within a de Sitter true vacuum background.  Although this
process is energetically forbidden in the absence of gravity, it can
occur in de Sitter spacetime with its nonzero temperature $T_{\rm
  dS}$~\cite{Lee:1987qc}.  In contrast with the CDL case, the bounces
governing the nucleation of true vacuum and false vacuum bubbles are
not identical.

The remainder of this paper is organized as follows.  In Sec.~II, we
review the essentials of vacuum tunneling, focusing on some features
of particular relevance here.  In Sec.~III, we set up the formalism
that we will use, and obtain some basic results concerning our
solutions.  In Sec.~IV we discuss the application of the thin-wall
approximation to our problem, explaining how its applicability is 
more limited than in the case of bounces with O(4) symmetry.  We then 
turn to the description of the ${\rm O(3)}\times{\rm O(2)}$-symmetric 
bounces themselves.  Some limiting cases that can be handled analytically 
are discussed in Sec.~V, and in Sec.~VI we describe the results of our
numerical solutions.  Section~VII contains some concluding remarks.

\section{Review of vacuum tunneling}

\subsection{Flat spacetime at zero temperature}

In flat spacetime at zero temperature, vacuum decay occurs through the
production of bubbles of true vacuum via quantum tunneling.
Importantly, this is not a tunneling process from homogeneous false
vacuum to homogeneous true vacuum, but rather tunneling from an
initial configuration that is pure false vacuum to a final
configuration containing a bubble of true vacuum (or an approximation
thereof) within a false vacuum background.  The tunneling is not
actually through the one-dimensional barrier that appears in a plot of
$V(\phi)$, but rather through an infinite-dimensional barrier in
configuration space defined by the potential energy functional
\begin{equation}
    U[\phi(x)] = \int d^3x \left[\frac12 ({\bm \nabla}_\mu \phi)^2 + V(\phi) 
             \right]   \, .
\end{equation}

One approach to this problem is via the WKB method.  With one degree
of freedom, $q(t)$, the WKB approximation gives an amplitude for
tunneling through a barrier that is proportional to $e^{-{\cal B}}$,
where ${\cal B}$ is of the form
\begin{equation}
    {\cal B} = 2 \int_{q_1}^{q_2} dq \sqrt{2[V(q)-E] }
\end{equation}
with $V(q_1)=V(q_2)=E$.  For a barrier in a multidimensional
configuration space, one must find a path through the barrier for
which the corresponding one-dimensional tunneling exponent ${\cal B}$
is a local minimum~\cite{Banks:1973ps,Banks:1974ij}.
Coleman~\cite{Coleman:1977py} showed that the problem of finding a 
stationary point of 
${\cal B}$ is equivalent to finding a ``bounce'' solution of the
Euclidean equations of motion.  In the field theory case, the bounce
is a solution $\phi({\bf x},\tau)$ in four-dimensional Euclidean
space.  The hypersurface at the initial value $\tau_{\rm init}$ (which
is equal to $-\infty$ for tunneling from the false vacuum) gives the
configuration before tunneling, while the configuration $\phi({\bf
  x},\tau=0)$ along the hypersurface through the center of the bounce
gives the optimal exit point from the potential energy barrier; i.e.,
the form of the bubble at nucleation.  The hypersurfaces at
intermediate Euclidean times $-\infty<\tau<0$ give successive
configurations along the optimal tunneling path.  Those for $0 < \tau
<\infty$ are simply mirror images of those for negative $\tau$, thus
motivating the term ``bounce''.  The bubble nucleation rate per unit
volume is of the form $\Gamma = A e^{-{\cal B}}$, with ${\cal B}$
being equal to the difference of the Euclidean actions of the bounce
and the pure false vacuum.

\begin{figure}[t]
\centering
\includegraphics[height=2.5in]{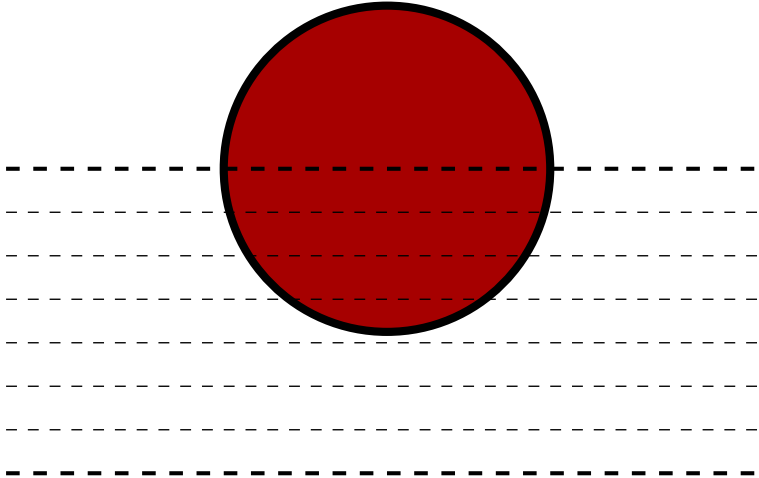}
\caption{Schematic view of a bounce in flat spacetime.  The shaded
  area is the region of approximately true vacuum, while the heavy
  circle surrounding it represents the wall region.  In actuality, the
  field is not strictly constant outside the wall, but instead has an
  exponential tail and only reaches its false vacuum value at infinite
  distance.  The horizontal dashed lines represent a sequence of
  spatial hypersurfaces of constant $\tau$ that trace out a path in
  configuration space.  The lower heavy dashed line (corresponding to
  $\tau = -\infty$) indicates the initial pure false vacuum state,
  while the heavy dashed line through the center of the bounce
  corresponds to the field configuration at bubble nucleation.}
\label{flat-bounce-sliced}
\end{figure}

Subject to only mild assumptions on the
potential~\cite{Coleman:1977th}, the bounce of minimum action can be
shown to have an O(4) symmetry.  This bounce has a central
region, in which $\phi$ is close to its true vacuum value, within an
infinite region of false vacuum, with the two regions separated by a
wall of finite thickness in which $\phi$ traverses the potential
barrier, as shown in Fig.~\ref{flat-bounce-sliced}.

An alternative approach is to use a Euclidean path integral to
calculate the imaginary part of the energy of the false vacuum, which
is closely related to $\Gamma$~\cite{Callan:1977pt}.  This gives the
prefactor $A$ in terms of the eigenvalues of small fluctuations about
the bounce, with the imaginary part of the energy emerging from the
presence of a single negative eigenvalue.  If there are additional
negative eigenvalues, the path specified by the bounce is not even a
local minimum of ${\cal B}$, but rather a saddle point, and the bounce
solution must be discarded~\cite{Coleman:1987rm}.

\subsection{Finite temperature}
\label{finiteT}

Both the WKB and the path integral approaches show that tunneling at a
nonzero temperature $T$ is governed by a bounce solution that is
periodic in imaginary time with period $\beta=1/T$.  Two types of
periodic bounces can be distinguished.   

At low temperature the bounce is a somewhat deformed and elongated
version of the zero-temperature bounce, with the O(4) symmetry reduced
to an O(3) symmetry.  Such bounces correspond to thermally assisted
tunneling processes in which thermal effects allow the system to start
the tunneling at a somewhat higher point on the potential energy
barrier.  At sufficiently high temperatures this solution may cease to
exist.

The second type of solution corresponds to a thermal fluctuation to a
saddle point at the top of a ``mountain pass'' over the potential
energy barrier.  These solutions are independent of $\tau$. Each
three-dimensional spatial slice is infinite in extent and describes an
O(3)-symmetric configuration $\phi({\bf x})$ with a central true
vacuum bubble of critical size.  The Euclidean action of these
solutions is equal to the energy of this critical bubble divided by the
temperature.  Although these solutions exist for all values of $T$, at
low temperature they have multiple negative modes and must therefore
be discarded.  Because this point will be important for us later, let
us examine it a bit more closely.

A critical bubble solution $\phi({\bf x})$ is a saddle point of
the potential energy functional $U[\phi]$.  The second variation of 
$U$ has a single negative mode $\psi_-({\bf x})$ satisfying 
\begin{equation}
      {\delta^2 U \over \delta \phi^2}\, \psi_- 
       = -{\bm \nabla^2} \psi_-({\bf x}) 
     + V''(\phi({\bf x}))\, \psi_-({\bf x}) = \lambda_- \psi_-({\bf x}) 
\end{equation}
with $\lambda_- = - k^2 < 0$.  The second variation of the Euclidean action,
\begin{equation}
  {\delta^2 S \over\delta \phi^2} = -{d^2 \over d\tau^2}
  -{\bm\nabla^2} + V''(\phi({\bf x}))  \, ,
\end{equation}
must then have a spectrum that includes eigenmodes of the form
\begin{equation}
  \cos \left({2\pi n \tau\over \beta}\right) \psi_-({\bf x})  \, \qquad n=0,1,\dots
\end{equation}
and 
\begin{equation}
    \sin \left({2\pi n \tau\over \beta}\right) \psi_-({\bf x})\, 
         \qquad n=1,2,\dots  
\end{equation}
with eigenvalues
\begin{equation}
   \lambda_n = -k^2 + \left({2\pi n \over \beta}\right)^2   \, .
\end{equation}
Hence, at temperatures $T < k/2\pi$ there are multiple negative modes,
and the $\tau$-independent bounce must be discarded.  In the flat
spacetime thin-wall approximation with a critical bubble of radius
$R$ one finds that $k= \sqrt{2}/R$.  More generally, we expect that $k
\sim 1/{\cal R}$, where ${\cal R}$ is a characteristic length scale of
the critical bubble solution.  Thus, we must require that
\begin{equation}
    T \gtrsim {1 \over 2\pi {\cal R}} \, .
\label{Rconstraint}
\end{equation}

\subsection{Including gravity}

Coleman and De Luccia  proposed~\cite{Coleman:1980aw} that the
effects of gravity on bubble nucleation could be obtained by including
a curvature term in the Euclidean action and seeking a bounce solution
that satisfies the coupled Euclidean Einstein and matter equations.
If one assumes that the bounce has an O(4) symmetry, and $V(\phi)$ is
everywhere positive, the bounce solution is topologically a
four-sphere that can be viewed as the hypersurface
\begin{equation}
     x_1^2 + x_2^2 +x_3^2 +x_4^2 +x_5^2  = H^{-2}
\label{fivespherecoords}
\end{equation}
in a five-dimensional Euclidean space, although not necessarily with
the metric implied by the embedding.   An O(4)-symmetric solution
is obtained by requiring invariance under rotations among the first
four coordinates.  The scalar field $\phi$ is then a function only of 
$x_5$.

\begin{figure}[t]
\centering
\includegraphics[height=2.0in]{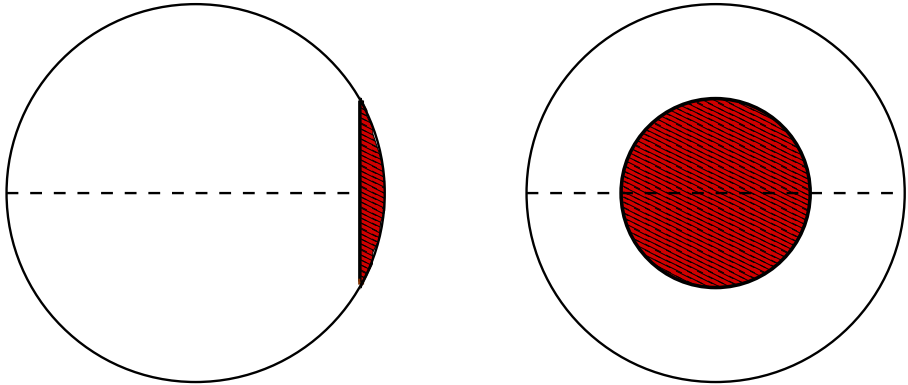}
\caption{Two views of the CDL bounce.  The bounce has the topology of a
four-sphere, and is shown here as if embedded in a five-dimensional
Euclidean space.  Because of the bounce's O(4) symmetry, the fields
and metric have nontrivial dependence only on $x_5$.  The left-hand
view shows the bounce as viewed from a point outside the bounce on an
axis perpendicular to the $x_4$-$x_5$ plane, where $x_4$ and $x_5$
correspond to the vertical and horizontal directions in the figure,
respectively.  The shaded area corresponds to the region of
approximately true vacuum.  The horizontal dashed line has often been
taken, by analogy with the heavy dashed line in
Fig.\ref{flat-bounce-sliced}, as giving the configuration at bubble
nucleation.  The right-hand view shows the bounce as viewed from a
point to its right on the positive $x_5$-axis.  The $x_4$ axis again
corresponds to the vertical direction, while the horizontal direction
now corresponds to an axis perpendicular to the $x_4$-$x_5$ plane..  A
view from a corresponding point on the negative $x_5$-axis would show
only a region of approximate false vacuum.}
\label{std-CDL}
\end{figure}

Figure~\ref{std-CDL} illustrates such a solution for the case where
the bubble nucleates with a size much smaller than the false vacuum
horizon length $H_{\rm fv}^{-1}$.  Analogy with the flat spacetime
case illustrated in Fig.~\ref{flat-bounce-sliced} might suggest that
the configuration after nucleation is given by a slice along the
``equator'' of this hypersphere that cuts through the center of the
true vacuum region, as shown in the figure, with the tunneling path
given by a sequence of parallel horizontal slices.  This
interpretation becomes problematic when the bubble size becomes
comparable to the horizon length and there is no plausible candidate
for the initial configuration.  The is especially so in the limit
where the bounce becomes the homogeneous Hawking-Moss
solution~\cite{Hawking:1981fz} and a straightforward application of
the interpretation would imply that the bounce corresponded to a
transition in which all of space fluctuated to $\phi_{\rm top}$, the
value at the top of the barrier in $V(\phi)$.

\begin{figure}[t]
\centering
\begin{tabular}{cc}
\includegraphics[height=2.0in]{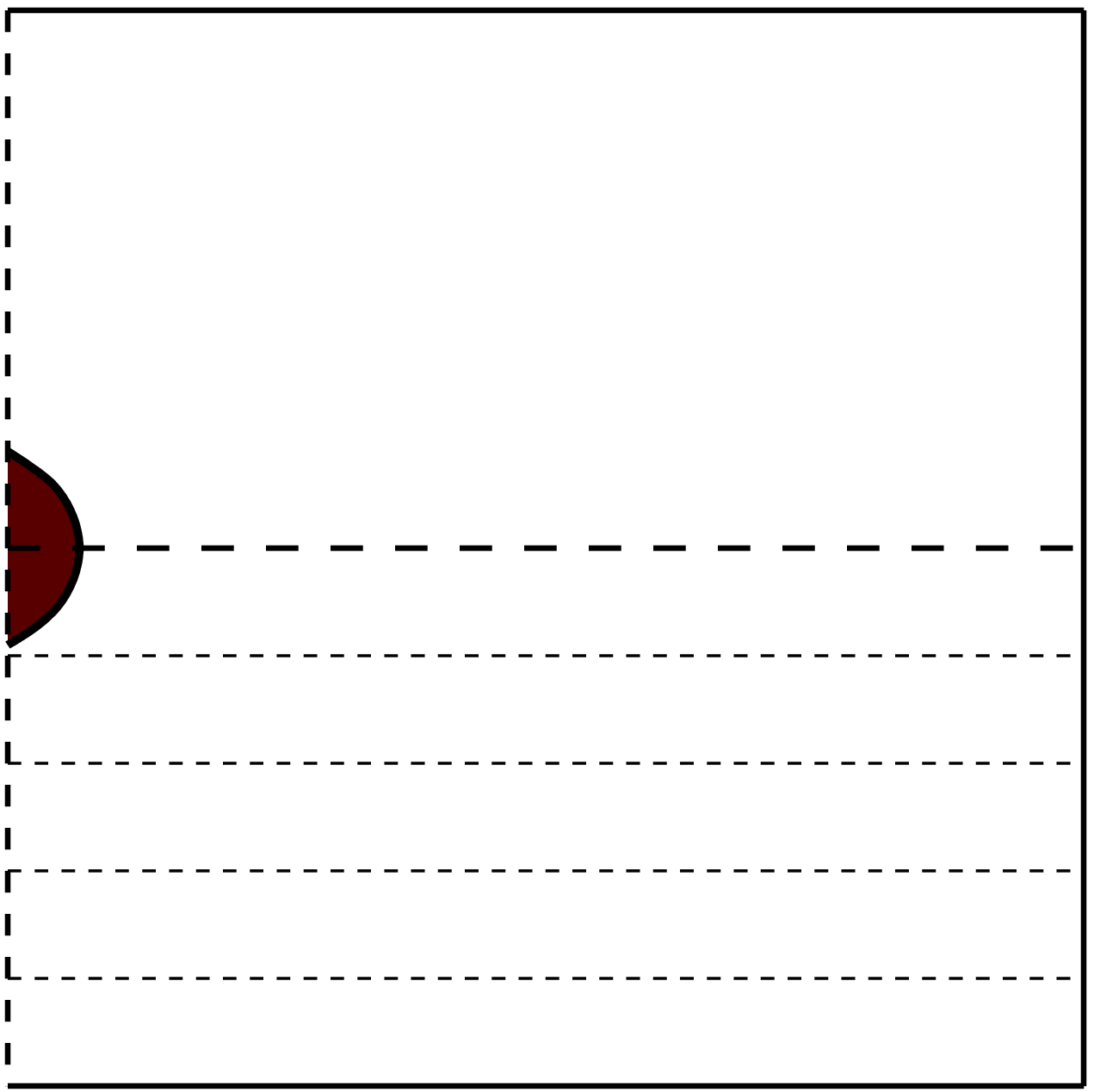}
\hskip .7in
\includegraphics[height=2.0in]{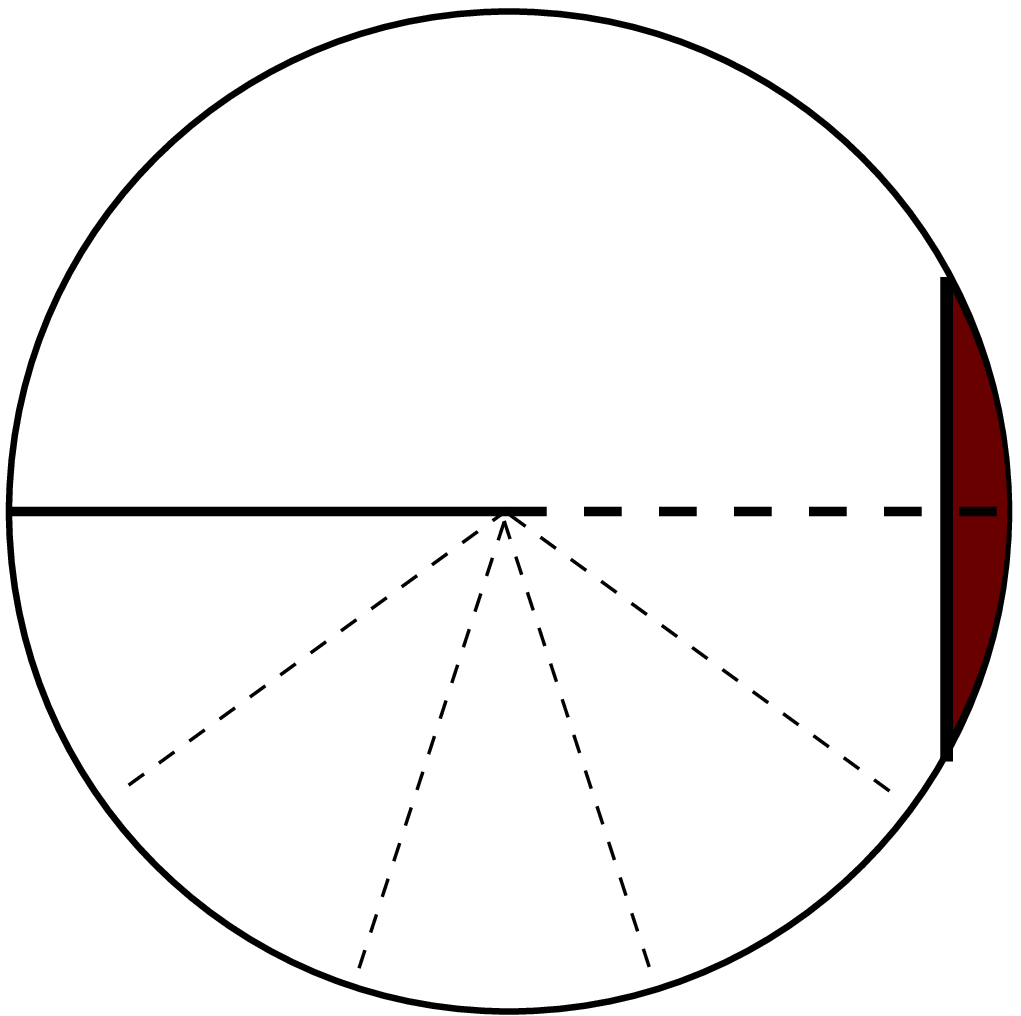}
\\(a) \hskip 2.5in (b) 
\end{tabular}
\caption{The CDL bounce viewed as low-temperature thermal tunneling
  within a horizon volume.  In (a) the horizontal lines represent
  three-balls with their center on the left and their outer edge, at
  the horizon, on the right.  The horizontal line at the bottom of the
  figure corresponds to the configuration before tunneling, while the
  one at the center of the figure corresponds to the configuration
  after tunneling.  As in Fig.~\ref{flat-bounce-sliced}, the
  intermediate horizontal lines define the optimum tunneling path
  through configuration space.  In (b) the same CDL solution is
  represented, but with the path through configuration space defined
  by a series of radial lines.  Again, each of these represents a
  three-ball, with center of the ball at the edge of the figure and
  the two-sphere boundary at the horizon at the center of the figure.
  The initial and final configurations correspond to the left solid
  and right dashed horizontal radii, respectively.}
\label{flat-sliced-CDL}
\end{figure}

In Ref.~\cite{Brown:2007sd} it was argued that one should instead
interpret the Coleman-De Luccia formalism as describing thermal
tunneling within a horizon volume, with the temperature being the one
defined by the horizon.  From this viewpoint, the solutions are
analogous to those in flat spacetime at finite temperature, but with
the crucial difference that the spatial slices of constant Euclidean
time are finite in extent.  Thus, these solutions exist on a space
that is topologically the product of a three-ball (the horizon volume)
and a circle (the periodic Euclidean time $\tau$).  This suggests a
``low-temperature'' bounce solution such as that shown in
Fig.~\ref{flat-sliced-CDL}a.  The boundary of this space is a
family of two-spheres, each labeled by a value of $\tau$.

At first sight, this seems nothing like the CDL bounce.  However, this
can also be represented as in Fig.~\ref{flat-sliced-CDL}b, with the
hypersurfaces of constant $\tau$ appearing as diagonal sections
cutting halfway through a four-sphere.  On each section the point at
the edge of the diagram corresponds to the center of the horizon
volume, while the bounding horizon radius two-sphere appears at the
center of the diagram, $x_4=x_5=0$.  Drawn in this manner, the bounce
looks like the four-sphere bounce of Fig.~\ref{std-CDL}.  It actually
{\it is} topologically a four-sphere if the two-sphere boundaries of
the sections are all identified.

There appears to be no a priori requirement that these two-spheres be
identified, nor that the metric should be smooth there if they are
identified.  Nevertheless, the O(4)-symmetric CDL bounce, with an
everywhere smooth metric, can be obtained in this fashion.  However,
we see that the equatorial slice is now composed of two parts.  One
half gives the field configuration at the time of bubble nucleation,
but only within the horizon volume.  The other half gives the field
configuration in this horizon volume {\it before} the tunneling
process.  Note that the bounce gives us no information about the field
or metric beyond the horizon, other than the fact that they
effectively create a thermal bath for the horizon volume.

With this understanding of the CDL bounce as describing
low-temperature thermally assisted tunneling, we see that the
gravitational analogue of the flat spacetime high-temperature bounce is
one that is independent of $\tau$, with the slices of constant $\tau$
describing the critical bubble configuration that is reached by
thermal fluctuation.  The Hawking-Moss solution is the simplest
example, and corresponds to a spatially homogeneous critical ``bubble''
that fills the horizon volume.  However, there can also be critical
bubbles that are smaller than horizon size.  One such is obtained
simply by rotating a CDL bounce so that its symmetry axis is, say, 
along the $x_1$ axis.  This rotated bounce corresponds to thermal
nucleation of a critical bubble centered at the
horizon~\cite{Brown:2007sd}.

\begin{figure}[t]
\centering
\includegraphics[height=2.0in]{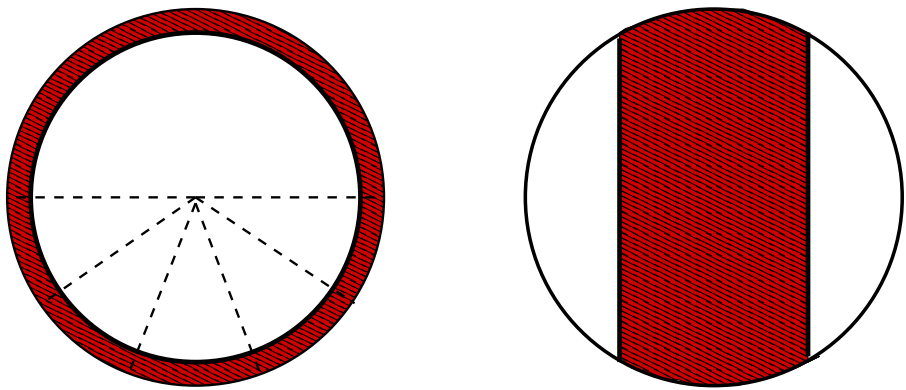}
\caption{Two views of the O(3)$\times$O(2)-symmetric bounce.  These
  are from the same viewpoints as those of the CDL bounce in
  Fig.~\ref{std-CDL}.  (The view on the right would be the same if
  viewed from along any axis in the $x_4$-$x_5$ plane.)  On the left-hand
  view the radial dashed lines correspond to the similar lines in
  Fig.~\ref{flat-sliced-CDL}b.  The bounce is independent of the
  Euclidean time, and the three-dimensional configuration corresponding
  to any one of these lines contains a single true vacuum critical
  bubble in the center of a false vacuum horizon volume.}
\label{o3o2-bounce}
\end{figure}

We will proceed by seeking a different, and perhaps more obvious, type
of bounce, namely one that is independent of $\tau$ and spherically
symmetric on each constant-$\tau$ slice.  This would correspond to
thermal production of a critical bubble located at the center of the
horizon volume.  Drawn as a four-sphere, the bounce would appear as in
Fig.~\ref{o3o2-bounce}.  We see here a case where the analysis of
Ref.~\cite{Brown:2007sd} leads to a prediction that is clearly
different from the conventional interpretation.  Whereas the latter
would view such bounces as describing the production of a pair of
bubbles, under the interpretation of Ref.~\cite{Brown:2007sd} only a
single bubble is produced.

\section{Formalism and some basic results}

We are thus led to seek solutions that have nontrivial behavior
in three spatial dimensions, but that are constant in Euclidean time. 
If we impose spherical symmetry on the spatial slices, the metric
on the four-dimensional space can be written in the form
\begin{equation}
   ds^2 = B(r)d\tau^2 + A(r) dr^2 +r^2 d\Omega^2
\label{o3o2metric}
\end{equation}
and the scalar field $\phi$ is a function only of $r$.  Each spatial
slice is a three-ball bounded by a two-sphere at $r=r_H$, with
the horizon radius $r_H$ corresponding to a zero of $g^{rr}=1/A$.  In order
that the metric be nonsingular at $r=0$, we must require that
$A(0)=1$.  We do not fix $B(0)$ directly, but instead require that it
be such that the product $AB$ is equal to unity at the horizon.
Generically, there is a region of approximate true vacuum centered
around the origin, $r=0$, an outer region of approximate false vacuum
extending out to the horizon, and a wall region separating the two.

In the analogous flat spacetime case the temperature, and hence the
periodicity, can be chosen arbitrarily, but in our curved spacetime
problem we expect the temperature to be specified by the metric.  We
therefore assume that the periodicity $\beta$ of the Euclidean time is
determined by the surface gravity at the horizon of the Lorentzian
counterpart of our Euclidean metric; i.e.,
\begin{equation}
    \beta = {1 \over T}
      = 4\pi \left[ B' \left({1\over A}\right)' \right]_{r_H}^{-1/2} \, ,
\label{betaEq}
\end{equation}
where a prime indicates differentiation with respect to $r$
and the subscript indicates that all quantities are to be evaluated
at the horizon.\footnote{Note that $\beta$ only enters our
  calculations as an overall multiplicative factor in the action, and
  does not otherwise affect the solutions.  It would be a
  straightforward matter to extend our results to an arbitrary
  temperature, although the physical origin of the temperature would
  be less compelling.}  Now note that
\begin{equation}
    B' = {d\over dr}\left[(AB) \left({1 \over A}\right)\right] 
    =(AB)' \,\left({1\over A}\right) + (AB)\, \left({1\over A}\right)' \, .
\end{equation}
Because $A^{-1} =0$ and $AB=1$ at the horizon, we have $B'=(A^{-1})'$ at the
horizon and 
\begin{equation}
    \beta = -4\pi [B'(r_H)]^{-1}  \, ,
\label{betaFormula}
\end{equation}
with the minus sign arising because $B'$ is negative at the horizon.  With a
uniform vacuum $\beta$ would be $2\pi r_H$; for the bounce solutions it remains
of this order of magnitude.

With this periodicity, the four-sphere
obtained by identifying all of the boundary two-spheres is a smooth
manifold, just as in the CDL solution.  However, as in the CDL case,
we view this as incidental and do not require $\phi$ to be smooth on
this two-sphere.  Such smoothness would require that $\phi'=d\phi/dr$
vanish at $r=r_H$. As we will see, this is only possible in the
(unattainable) limit in which the thin-wall approximation is exact.
Although having $\phi'(r_H) \ne 0$ implies a discontinuity in slope
along a path passing through the two-sphere, this discontinuity does not imply any
cost in action, and we see no problem with this.

The Euclidean action is 
\begin{eqnarray}
    S &=&\int d^4x \sqrt{g} 
      \left( -{1 \over 16\pi G}R + {\cal L}_{\rm matter} \right)  \cr
     &=& 4\pi\beta \int_0^{r_H} dr \, r^2 \sqrt{AB}
       \left[ -{1 \over 16\pi G}R
        + \frac12 {(\phi')^2 \over A} + V(\phi) \right] \, .
\end{eqnarray}
The curvature scalar that follows from the metric of Eq.~(\ref{o3o2metric}) is
\begin{equation}
   R = - {1 \over r^2 \sqrt{AB}}{d\over dr}\left[r^2 {B' \over \sqrt{AB}} \right]
          + {2\over r^2} {d\over dr} \left[r \left(1 -\frac1A\right)\right] \, .
\end{equation}
Substituting this back into the action gives
\begin{equation} 
  S = {\beta \over 4G} \left[r^2 {B' \over \sqrt{AB}} \right]_{r=0}^{r=r_H}
+ 4\pi\beta \int_0^{r_H} dr \,  \sqrt{AB} \left\{
     {1 \over 8\pi G} {d\over dr} \left[r \left(\frac1A -1\right)\right]
          +r^2 \left[\frac12 {(\phi')^2 \over A} + V(\phi)\right] \right\} \, .
\label{actionFormula}
\end{equation}

Although this is written in terms of the three functions $A(r)$, $B(r)$, and 
$\phi(r)$, it is more convenient to treat $AB$, rather than $B$, as the independent
variable.  Varying the action with respect to the independent variables then 
leads to\footnote{These equations can also be obtained by substituting our ansatz for the
field and metric into the full field equations.}
\begin{eqnarray}
  0 &=& {1 \over 8\pi G}{d\over dr} \left[r \left({1 \over A} -1\right)\right]
    +r^2 \left[\frac12 {(\phi')^2 \over A} + V(\phi) \right] \, ,
               \label{Aeq} \\ \cr
  0 &=& {(\sqrt{ AB})' \over \sqrt{AB}} 
      - 4  \pi G r (\phi')^2  \, , \label{ABeq}   \\ \cr 
  0 &=& { 1 \over r^2 \sqrt{AB}} {d\over dr} 
         \left[ r^2 \sqrt{AB} \,  {\phi'\over A} \right] - {dV \over d\phi} \, .
        \label{phieq}
\end{eqnarray}
For any solution of these equations [or even any configuration
  satisfying Eq.~(\ref{Aeq})], the integral in
Eq.(\ref{actionFormula}) vanishes.  Evaluating the first term with the
aid of Eq.~(\ref{betaFormula}), and recalling our convention that
$AB=1$ at the horizon, we obtain
\begin{equation}
     S_{\rm bounce} = - {\pi \over G}\, r_H^2 \, .
\end{equation}
This agrees with the thin-wall result of
Ref.~\cite{Garriga:2004nm}, and could in fact have been anticipated
on more general grounds~\cite{Banados:1993qp,Hawking:1995fd}.

Finally, the exponent in the tunneling rate is the difference between the 
Euclidean actions of the bounce and of the original false vacuum; i.e.,
\begin{equation}
   {\cal B} = S_{\rm bounce} - S_{\rm fv} 
         = {\pi \over G} ( \Lambda^2_{\rm fv} -r_H^2 )  \, ,
\label{BforUs}
\end{equation}
where the false vacuum horizon radius
\begin{equation}
   \Lambda_{\rm fv} = H^{-1}_{\rm fv} = \sqrt{3\over 8\pi GV_{\rm fv}} \, .
\end{equation}

For actually seeking bounce solutions, it is convenient to recast our
field equations a bit.  First, the equations are simplified by defining
\begin{equation}
      f = \frac1A   \, .
\end{equation}
Equation~(\ref{Aeq}) then becomes
\begin{equation}
   0 = {1\over 8\pi G} {d\over dr}\left[ r(f-1) \right]
         +r^2\left[ \frac12 (\phi')^2f + V(\phi) \right] \, .
\label{Aeq2}
\end{equation}
If we define  a mass ${\cal M}(r)$ by
\begin{equation}
  f(r) = 1 - {2G {\cal M}(r) \over r}  \, ,
\label{massdef}
\end{equation}
this can be rewritten as 
\begin{equation}
     {\cal M}' = 4\pi r^2 \left[\frac12  (\phi')^2 f  + V(\phi) \right] \, .
\label{masseq}
\end{equation}

Second, we note that substitution of Eq.~(\ref{ABeq}) into
Eq.~(\ref{phieq}) gives
\begin{equation}
  0 = f\left[\phi'' +  \frac2r \, \phi'+ 4\pi G r  (\phi')^3 \right]
         + f' \phi' - {dV \over d\phi}  \, .
\label{phiEq2}
\end{equation}
This and Eq.~(\ref{masseq}) give a pair of equations involving only
$f$ and $\phi$.  Once these have been solved, $AB$ is readily obtained
by integrating Eq.~(\ref{ABeq}).  With our convention that $AB$ be
equal to unity at the horizon, this gives
\begin{equation}
   A(r)B(r) = \exp\left[-8\pi G\int_r^{r_H} dr'\, r' \, (\phi')^2 \right] \, .
\label{ABsolution}
\end{equation}

We must now determine the boundary conditions.  We clearly need to avoid
singularities at the center of the bounce, $r=0$.  This leads to the
requirements 
\begin{eqnarray}
    f(0) &=&  1  \, , \cr
    \phi'(0) &=& 0  \, .
\end{eqnarray}
The value of $\phi(0)$ is not predetermined, although it must be 
on the side of the barrier corresponding to the new vacuum (i.e., 
the true vacuum side for the case of false vacuum decay by bubble 
nucleation).  

The horizon is determined by the requirement that $f(r_H)=0$.  Our
prescription for specifying the temperature guarantees that the
Euclidean metric is nonsingular at $r_H$, even when the bounce is
viewed as a topological four-sphere and not simply as the product of a
three-ball and a circle.  In order for $\phi$ to also be nonsingular
on the four-sphere we would have to require that $\phi'(r_H) =0$.
Equation~(\ref{phiEq2}) shows that this implies that $dV/d\phi =0$ at
the horizon, which in turn means that the field must be precisely at a
vacuum value (or else exactly at the top of the barrier, as in the
Hawking-Moss solution).  For a nontrivial bounce this is only possible
in the limit where the thin-wall approximation is exact.

However, there is no reason to require such smoothness on the four-sphere.  
Instead, the only requirement is that following from 
Eq.~(\ref{phiEq2}), which implies that
\begin{equation}
   0 = \left.\left[ f'\phi' -{dV \over d\phi}\right]\right|_{r=r_H}
\end{equation}
if $\phi'$ and $\phi''$ remain finite as $r\rightarrow r_H$.  [In fact,
the condition holds even for singular $\phi'$, provided that $\phi'$
grows more slowly than $(r_H-r)^{1/2}$ as the horizon is approached.
Note that such a divergence does not imply a divergent action density,
and in fact is consistent with a finite derivative of $\phi$ with
respect to proper distance from the horizon.]

\section{The thin-wall approximation}

The thin-wall approximation is often useful for gaining an intuitive
understanding of tunneling problems.  In this approximation, the
bounce is approximated as a region of pure true vacuum surrounded by a
region of pure false vacuum, with the variation of the scalar field
confined to a thin wall separating the two.  There are two essential
requirements for this approximation to be valid.  First, the
thickness of the wall must be much less than the radius of the true
vacuum region, so that the wall can be locally approximated as being
planar.  The second concerns the position dependence of the surface
tension $\sigma$ of the wall.  Although $\sigma$ was independent of
position in most previous tunneling calculations, position dependence
is possible.  However, for the thin-wall approximation to be reliable,
the variation of $\sigma$ must be small on distance scales comparable to 
the width of the wall.

In flat spacetime the thickness of the wall is determined primarily by
the form of the potential barrier separating the true and false vacua.
Reducing  $\epsilon$, the difference in the energy densities of the
two vacua, while keeping the shape of the potential barrier otherwise
fixed increases the radius of the true vacuum region.  By making
$\epsilon$ sufficiently small one can always reach the regime where
the thin-wall approximation is valid.  This is the case both for
O(4)-symmetric zero-temperature bounces and for high-temperature
solutions that describe a critical bubble in three dimensions.

With gravity included, the situation is more complicated.  Because the
false vacuum horizon length limits the size of the bounce, the true
vacuum region cannot be made arbitrarily large.  To even have a
possibility of the thin-wall approximation being valid, the shape of
the potential barrier must be such that the natural width of the wall
is small compared to the horizon length.  If this is the case, then an
appropriate reduction of $\epsilon$ can ensure the validity of the
approximation for the O(4)-symmetric bounces.  

This turns out to not be true for our O(3)$\times$O(2)-symmetric
bounces.  To understand this, let us begin by assuming that the
variations in the metric functions $f$ and $B$ as one passes through
the wall are small enough that they can be treated as constants there.
Equation~(\ref{phiEq2}) then reduces to
\begin{equation}
    0 = f \left( \phi'' +\frac2r \phi' \right)
           - {dV \over d\phi}  \, .
\end{equation}
If $f$ is treated as a constant near the wall, this is identical with
the analogous flat spacetime equation, except for a rescaling $r
\rightarrow \tilde r = r /\sqrt{f}$.  In other words, the wall
thickness, measured in terms of $r$, is reduced by a factor of
$\sqrt{f}$ compared to the flat spacetime case; this is to be expected,
since it is $\tilde r$ that measures the proper distance across the
wall.  Now let us further assume that the second, $2\phi'/r$, term can
be neglected.  Multiplying the above equation by $\phi'$ then leads to
\begin{equation}
   0 = {d \over dr} \left[\frac12 (\phi')^2f - V(\phi) \right]
\end{equation}
so that the quantity in brackets is constant through the wall.  Evaluating
this constant just outside the wall tells us that in the wall 
\begin{equation}
    \sqrt{f} \, \phi' = \sqrt{2[V(\phi) -V_{\rm fv}]}  \, .
\end{equation}
Furthermore, we have
\begin{eqnarray}
   \int_{\rm wall} dr\, {f \over 2} (\phi')^2 
    &=& \int_{\rm wall} dr \, [V(\phi) -V_{\rm fv}]  
    = \int_{\phi_{\rm tv}}^{\phi_{\rm fv}} {d\phi \over \phi'} \, 
          {f \over 2} (\phi')^2   \cr
    &=&  {\sqrt{f} \over 2}\int_{\phi_{\rm tv}}^{\phi_{\rm fv}} 
           d\phi \sqrt{2[V(\phi) -V_{\rm fv}]}  
\end{eqnarray}
where the integrals over $r$ are restricted to the wall region.  

We can therefore define a surface tension
\begin{equation}
    \sigma = \int_{\phi_{\rm tv}}^{\phi_{\rm fv}}
           d\phi \sqrt{2[V(\phi) -V_{\rm fv}]}
\label{sigmaDef}
\end{equation}
and write the contribution to the action from the matter in the wall as
\begin{equation}
    4\pi \beta \int_{\rm wall} dr\, r^2 \sqrt{AB}
        \left[{f\over 2}(\phi')^2 + V(\phi) - V_{\rm fv}\right]
    = 4 \pi R^2 \beta \sqrt{f}\sqrt{AB} \,\sigma  \, .
\label{wallAction}
\end{equation}
After the gravitational contribution to the wall action is included,
Eq.~(\ref{wallAction}) agrees with the corresponding expression in
Ref.~\cite{Garriga:2004nm}, verifying that the surface tension that we
have defined in Eq.~(\ref{sigmaDef}) corresponds to that used in the
latter paper.

We must now go back and examine the validity of the assumptions that we
have made.  In particular, we must show that the fractional changes
in $r$, $f$, and $AB$ as one goes through the wall are all small.
For the first of these, we note that the thickness of the wall is
\begin{equation}
    \Delta r = \sqrt{f}\int_{\phi_{\rm tv}}^{\phi_{\rm fv}} 
         {d\phi \over \sqrt{2[V(\phi) -V_{\rm fv}]}}  \, .
\end{equation}
Just as in flat spacetime, this can be made arbitrarily small, without
materially changing $\sigma$, by making the potential barrier higher
by a factor $\gamma^2>1$ and narrower by a factor $1/\gamma$.  Also, for
fixed barrier height and width this expression decreases as one
moves toward the horizon.

Equation~(\ref{ABsolution}) shows that the fractional change in $AB$ is 
\begin{equation}
      { \Delta(AB) \over AB } =  8\pi G r \int_{\rm wall} dr 
    \, { (\phi')^2 } = 8\pi G r\sigma / \sqrt{f} \, .
\end{equation}
From Eqs.~(\ref{massdef}) and (\ref{masseq}) we find that
\begin{eqnarray}
  \Delta f &=& -{2G \over r} \Delta {\cal M}
  = -{8 \pi G r} \int_{\rm wall} dr \left[\frac12 f (\phi')^2 
       + V \right] \cr
   &=& - 8 \pi G r \left( \sigma \sqrt{f} + V_{\rm fv} \Delta r
         \right)  \, .
\end{eqnarray} 
Requiring that the fractional changes in $AB$ and in $f$ 
be small gives the constraint,
\begin{equation}
    8 \pi G r \sigma  \ll \sqrt{f} < 1  \, .
\label{sigmalimit}
\end{equation}
This will certainly fail if the bubble wall is too close to the
horizon, where $f$ vanishes.\footnote{On the other hand, if $r$ is
  much smaller than $r_H$, Eq.~(\ref{Rconstraint}) will be violated,
  in which case we expect to have multiple negative modes and
  therefore an unacceptable bounce.}  In particular, the limiting
surface tension $\sigma_N$ that is discussed in
Ref.~\cite{Garriga:2004nm} does not satisfy Eq.~(\ref{sigmalimit}).
In the field theory context, if the potential is such that
Eq.~(\ref{sigmaDef}) implies $\sigma \gtrsim \sigma_N$, then the
thin-wall approximation is not valid.  

\section{Limiting cases}
\label{limit-sec}

Before we turn to our numerical results, it may be helpful to 
consider some limiting cases.

\subsection{Weak gravity}

Let us begin by focusing on the effect of varying $G$, with all other
parameters held fixed.  We first consider the case where $G$ tends
toward zero, so that gravitational effects would be expected to
be small.  If $G$ is sufficiently small, $\phi$ varies from a value
near the true vacuum to one exponentially close to $\phi_{\rm fv}$ in
an interval $0 \le r \le \tilde r$, where $\tilde r$ is much less than
$\Lambda_{\rm fv}$.
In this region Eq.~(\ref{phiEq2}) is well approximated by the corresponding
flat spacetime equation.  Outside this region,
integration of Eq.~(\ref{masseq}) gives
\begin{equation}
  {\cal M}(r) = E_{\rm flat} + {1\over 2G} \, {r^3 \over \Lambda_{\rm fv}^2} \, ,
\end{equation}
where $E_{\rm flat}$ is the energy of the critical bubble in flat
spacetime.  To leading order
the horizon, where $f$ vanishes, is at
$r_H = \Lambda_{\rm fv} - G E_{\rm flat}$.
The tunneling exponent, Eq.~(\ref{BforUs}), is given by  
\begin{equation}
   {\cal B} = 2\pi \Lambda_{\rm fv} E_{\rm flat} 
            = E_{\rm flat}/T_{\rm dS}   \, ,
\end{equation}
where $T_{\rm dS} = 1/2\pi \Lambda_{\rm fv}$ is the de Sitter
temperature of the false vacuum.  As expected, this is the Boltzmann
exponent one would obtain for nucleation of a critical bubble in flat
spacetime at a temperature $T= T_{\rm dS}$.  However, because $T \ll 1
/\tilde r$, the high temperature bounce has multiple negative
modes, and must therefore be discarded.

\subsection{Strong gravity}

Now consider the opposite limit, with $G$ becoming large while all
other parameters are held fixed.  In this limit $\Lambda_{\rm fv}$
decreases, leaving less and less space for the wall separating the two
vacua.  As $\Lambda_{\rm fv}$ becomes smaller than the natural width
of the wall, $\phi$ is restricted to an increasingly narrow range of
values around the top of the barrier.  Eventually, the solution
becomes spatially homogeneous, with $\phi=\phi_{\rm top}$ everywhere.
This is, in fact, the usual Hawking-Moss solution, whose O(5) symmetry
has both O(4) and O(3)$\times$O(2) symmetries as subgroups.

The approach to the Hawking-Moss solution can be studied by taking
$\delta \phi \equiv \phi_{\rm top} - \phi $ to be small.  This
quantity enters quadratically in the equations for the metric
functions, Eqs.(\ref{Aeq}) and (\ref{ABeq}), so to leading order we
can linearize the scalar field Eq.~(\ref{phieq}) in $\delta
\phi$ and take the metric functions to be those of a de Sitter
spacetime with $V= V(\phi_{\rm top})$. The analysis is then quite
analogous to the four-dimensional case that was considered in
Ref.~\cite{Hackworth:2004xb}, with the main difference being the
replacement of Gegenbauer functions by Legendre functions.  We find
that the nontrivial bounce solution goes over to the Hawking-Moss
solution when\footnote{In~\cite{Hackworth:2004xb} it was found that
  there is on O(4)-symmetric solution merging with the Hawking-Moss
  solution whenever $V''/H^2$ at the top of the barrier is equal to
  $N(N+3)$ for any integer $N$.  For the O(3)$\times$O(2)-symmetric
  solutions the corresponding critical values are $2(N+1)(2N+1)-2$.}
\begin{equation}
    {V''(\phi_{\rm top}) \over H_{\rm top}^2} \le 10   \, .
\label{HMcondition}
\end{equation}

\subsection{Large vacuum energy}

Another possibility is to keep $G$ fixed, but to raise the vacuum
energies by adding a large constant to $V(\phi)$.  On a pure vacuum
solution increasing the vacuum energy has the same effect on the
metric as increasing $G$.  We would expect the same to be true if the
variation of $V(\phi)$ in the region between the two vacua is small
compared to the absolute magnitude of the potential.  As in the previous case,
the bounce solution should go over to the Hawking-Moss solution
when Eq.~(\ref{HMcondition}) is satisfied.

\section{Numerical results}
\label{numeric-sec}

To explore the properties of the bounces more closely, we considered a
theory with a quartic potential with two minima.  By shifting the zero
of $\phi$ so that the minima are equally spaced about the origin, with
$\phi_{\rm tv}= v$ and $\phi_{\rm fv} = -v$, any such potential can be
brought into the form
\begin{equation}
    V(\phi) = \lambda \left(C_0 v^4 - k v^3 \phi -\frac12 v^2 \phi^2
         +{k\over 3} v \phi^3 + \frac14 \phi^4\right)   \, .
\end{equation}
with $0<k<1$.  The top of the barrier separating the two vacua is at
$\phi = -kv$.  The theory is thus characterized by four dimensionless
quantities: $\lambda$, $C_0$, $k$, and
\begin{equation}
   h = 8 \pi G v^2   \, .
\end{equation}

The dependence on one of these, $\lambda$, is rather simple.  If
$\phi(x) = g(x)$ is a bounce solution for a given value of $\lambda$,
then $\phi(x) = g(\gamma x)$ is a solution for the theory with
$\lambda$ replaced by $\gamma^2\lambda$.  The action of the bounce in
the latter theory is $\gamma^{-2}$ times that of the original bounce.
Thus, the classical solutions have a nontrivial dependence on only the
three remaining dimensionless parameters.  At the same time, it must
be remembered that the validity of the semiclassical method requires
that we be in the weak coupling regime $\lambda \ll 1$.

We will find it useful to work with an alternative set of parameters,
\begin{eqnarray}
    \epsilon &=& {1\over  v^4} (V_{\rm fv}-V_{\rm tv}) 
        = \frac43 \lambda k \, , \cr\cr
   \alpha &=&  {1\over  v^4} (V_{\rm top}-V_{\rm fv})  
       = {\lambda\over 12}(3-k)(1+k)^3 \, , \cr \cr
   U_0 &=& {1\over v^4} V_{\rm fv}  
   = \lambda\left(-\frac12 +{2k \over 3} + C_0\right) \, ,
\end{eqnarray}
that characterize the splitting of the vacuum energies, the height of the
potential barrier, and the overall magnitude of the vacuum energy.  In
order to understand the dependence on these features of the potential,
we will hold two of these (and $h$) fixed while varying the third.
Such a variation corresponds to a more complicated path in the
$\lambda$-$C_0$-$k$ space, and may extend into the large $\lambda$
strong coupling regime.  This does not matter for our purposes, since
any such strong-coupling solution is related to a weak-coupling solution
whose field and metric profiles differ from the original only by a 
rescaling of distance

\begin{figure}
\centering
\begin{tabular}{cc}
\includegraphics[height=2.0in]{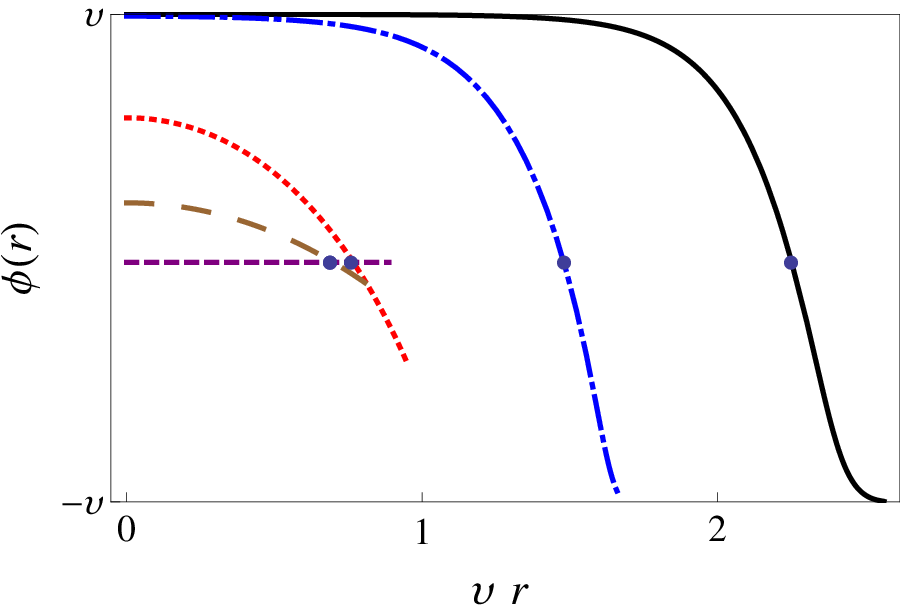}
\hskip .5in
\includegraphics[height=2.0in]{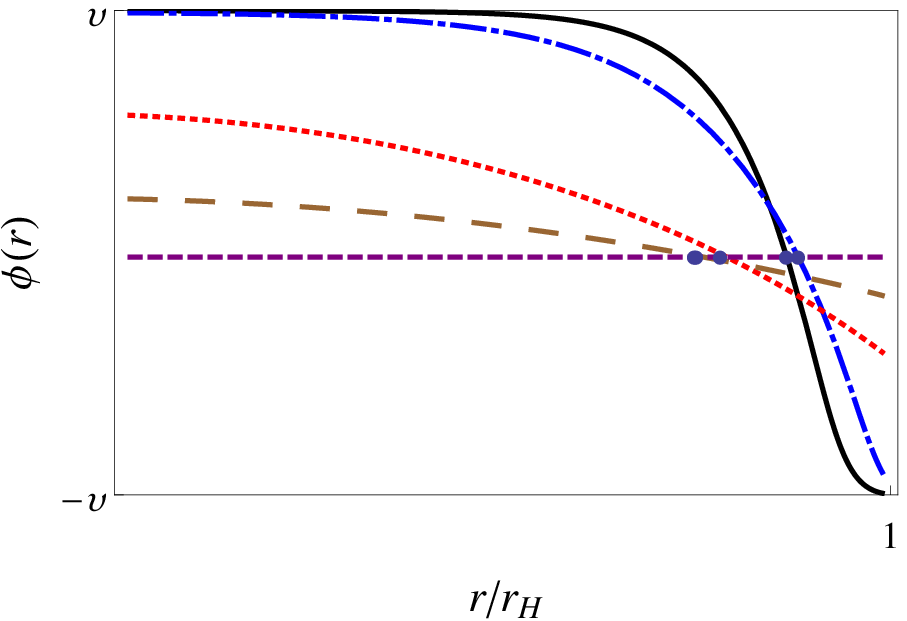}
\\(a) \hskip 3.0in (b) \\
\includegraphics[height=2.0in]{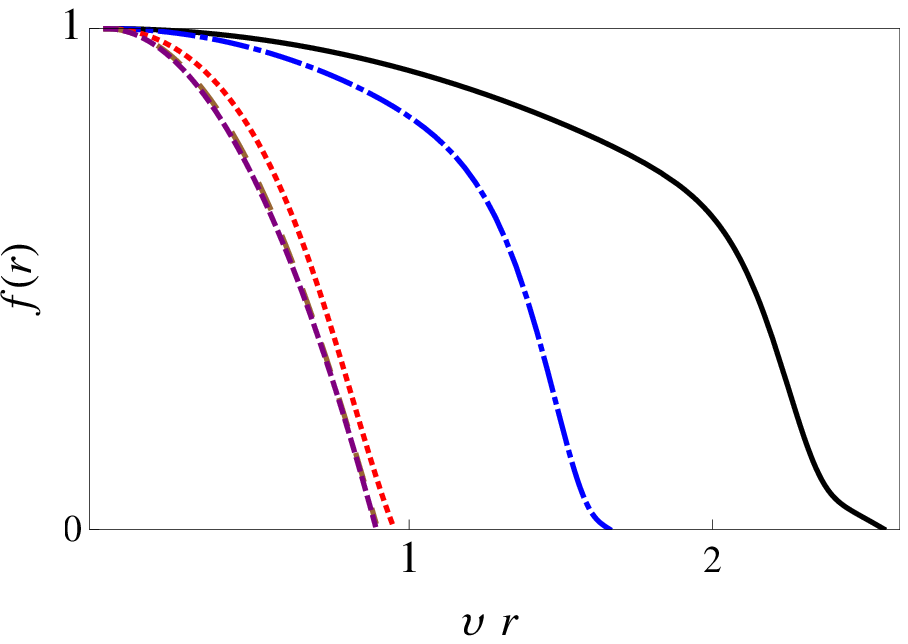}
\\  (c)
\end{tabular}
\caption{Evolution of the bounce as the gravitational constant is varied.
The first two panels show $\phi$ as a function of (a) $r$ and (b) $r/r_H$.  In 
both cases the black dot indicates the value at the top of the barrier.
The third panel shows $f=1/A$ as a function of $r$.  Reading from left to right, 
the short-dashed purple, long-dashed brown, dotted red, dot-dashed blue, and 
solid black lines correspond to $8\pi G v^2$ equal to 0.711, 0.704, 0.628, 0.251,
and 0.126.  In all cases $U_0=2$, $\alpha=3$, and $\epsilon=0.3$.}
\label{varyG}
\end{figure}

We have numerically solved the bounce equations for a variety of
values of the parameters.  Figure~\ref{varyG} shows the effect of
varying $G$ with the other parameters held fixed.  The horizon radius
decreases as gravity is made stronger, just as one would expect.  For
sufficiently strong gravity, the starting and ending points of the
field move significantly away from their respective vacuum values.
The solution eventually approaches the Hawking-Moss solution, with
$\phi$ being constant and $f$ precisely of the de Sitter form, in
agreement with the arguments that we presented in
Sec.~\ref{limit-sec}.  The Hawking-Moss solution is reached at the
value predicted by Eq.~(\ref{HMcondition}).

We also argued in Sec.~\ref{limit-sec} that the effect of uniformly
increasing $V_{\rm fv}$ should be qualitatively similar to that of
increasing the strength of gravity.  This is verified by our numerical
solutions, which closely resemble those shown in
Fig.~\ref{varyG}.  Again, the Hawking-Moss solution is reached at the
value predicted by Eq.~(\ref{HMcondition}).

\begin{figure}
\centering
\begin{tabular}{cc}
\includegraphics[height=2.0in]{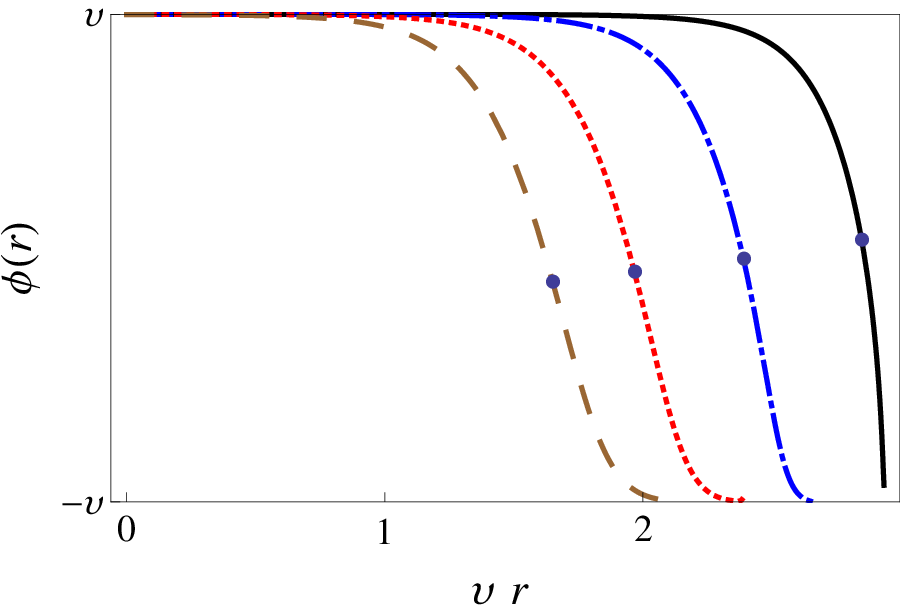}
\hskip .5in
\includegraphics[height=2.0in]{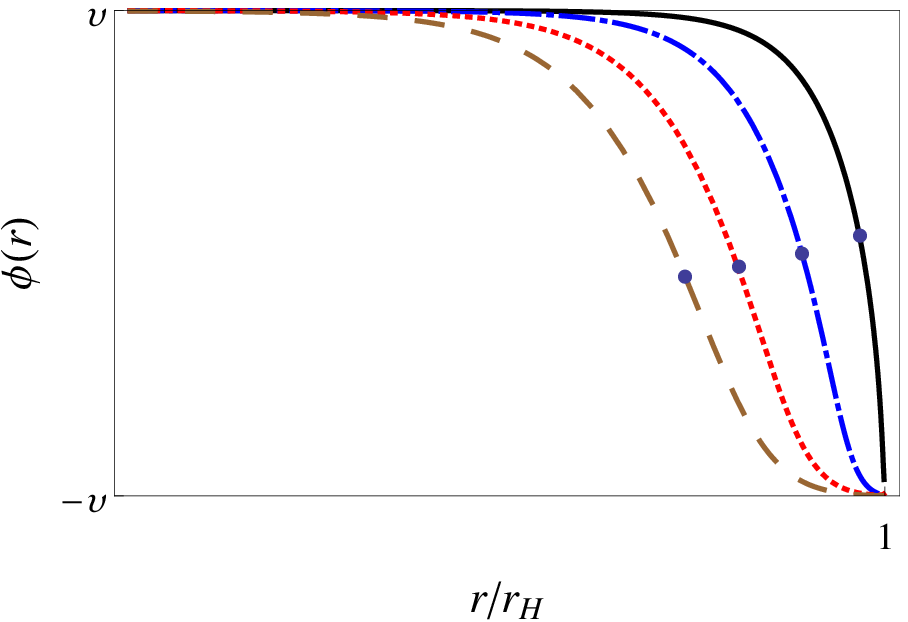}
\\   (a) \hskip 3.0in (b) \\
\includegraphics[height=2.0in]{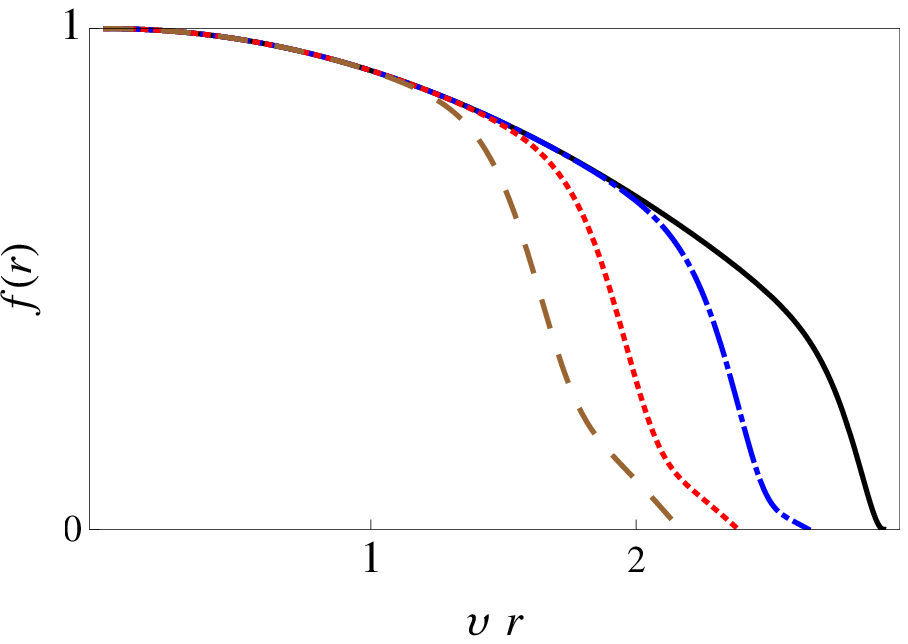}
\\   (c)
\end{tabular}
\caption{Behavior of the bounce as $\epsilon$ is varied.  Again, 
  panels (a) and (b) show $\phi$ as a function of $r$ and of $r/r_H$,
  with the black dot indicating the value at the top of the barrier.
  Panel (c) shows the metric function $f=1/A$.  Reading from left
  to right, the dashed brown, dotted red, dot-dashed blue, and solid
  black lines correspond to $\epsilon$ equal to 2, 1, 0, and $-1$.
  For all cases $U_0=2$, $\alpha=3$, and $8\pi G v^2 = 0.126$.}
\label{varyEps}
\end{figure}

Figure~\ref{varyEps} shows the effect of varying $\epsilon$, the
dimensionless difference of the two vacuum energy densities.  We see
that decreasing $\epsilon$ increases the bubble radius.  This behavior
is already familiar from the flat spacetime case, where the increased
radius is needed for the outward pressure of the interior true vacuum
to be able to overcome the surface tension of the bubble wall.  Note,
though, that we have included a curve for $\epsilon=0$, corresponding
to degenerate vacua, and that the outermost curve is for negative
$\epsilon$; i.e., the case of a false vacuum bubble forming within a
true vacuum region.  Neither of these would be possible in flat
spacetime, where the bubble radius tends to infinity as $\epsilon$ is
taken to zero, and where a true vacuum region cannot nucleate false
vacuum bubbles.  This is not so when gravitational effects are
included.  If the true and false vacua are both de Sitter, it is
possible to ``tunnel upward'' from true vacuum to
false~\cite{Lee:1987qc}.  Although the surface tension and the vacuum
pressure would both tend to collapse the resulting bubble, the Hubble
flow of de Sitter spacetime overcomes these if the initial bubble size is
sufficiently large.

In the CDL ansatz, the ``new vacuum'' region of the bounce is centered
about the ``North Pole'' of the four-sphere, and the ``old vacuum''
region about the ``South Pole''.  Nucleation of a bubble of false
vacuum rather than true is described by the same bounce solution, but
with the labeling of the two poles simply interchanged.  Matters are
different for the O(3)$\times$O(2)-symmetric bubbles.  The spacelike
slices of these are three-balls, with the new vacuum near the center
and the old vacuum at the outer edge.  There is no longer a symmetry
between the two regions, and the bounce profiles for the two processes
differ, as can be seen in the figure.

\begin{figure}
\centering
\begin{tabular}{cc}
\includegraphics[height=2.0in]{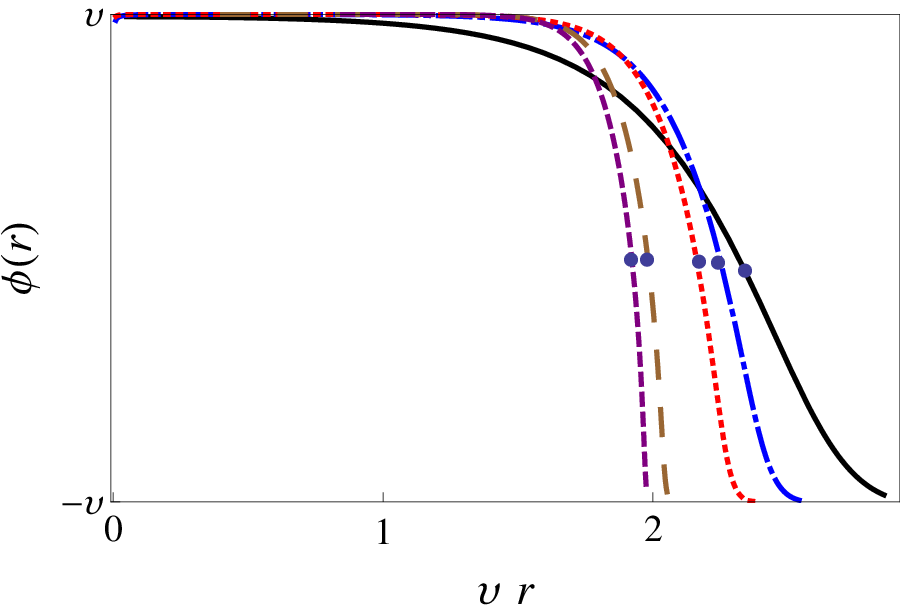}
\hskip .5in
\includegraphics[height=2.0in]{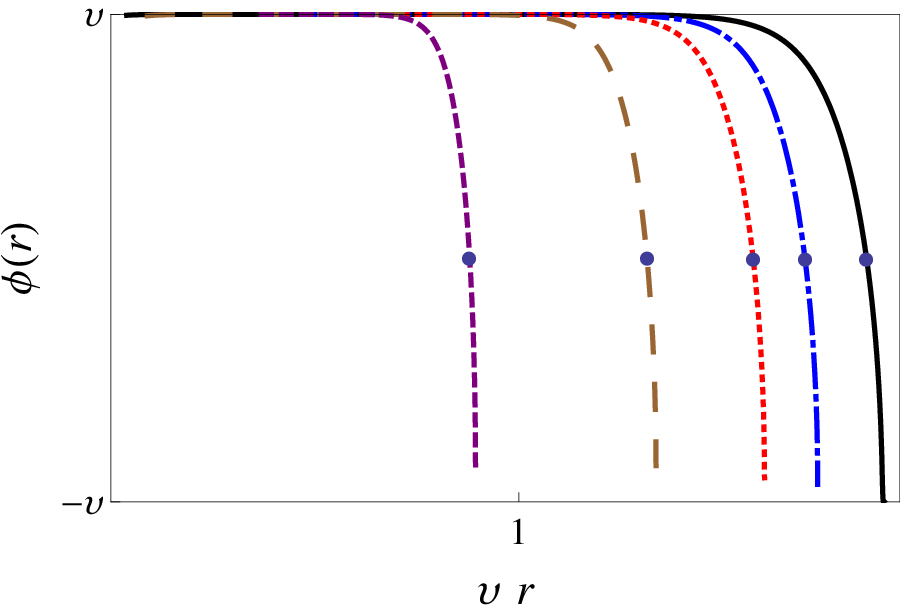}
\\   (a) \hskip 3.0in (b) \\
\includegraphics[height=2.0in]{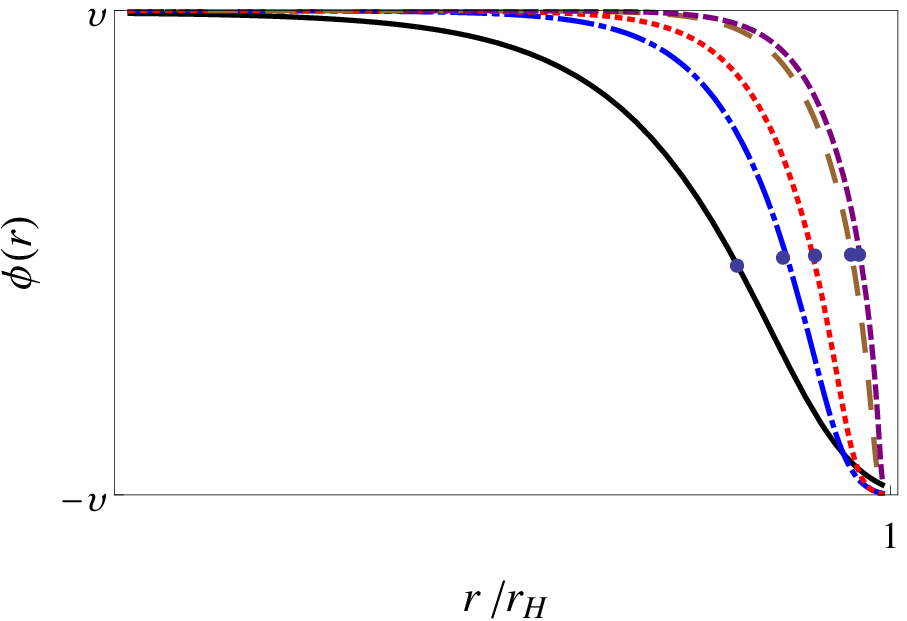}
\hskip .5in
\includegraphics[height=2.0in]{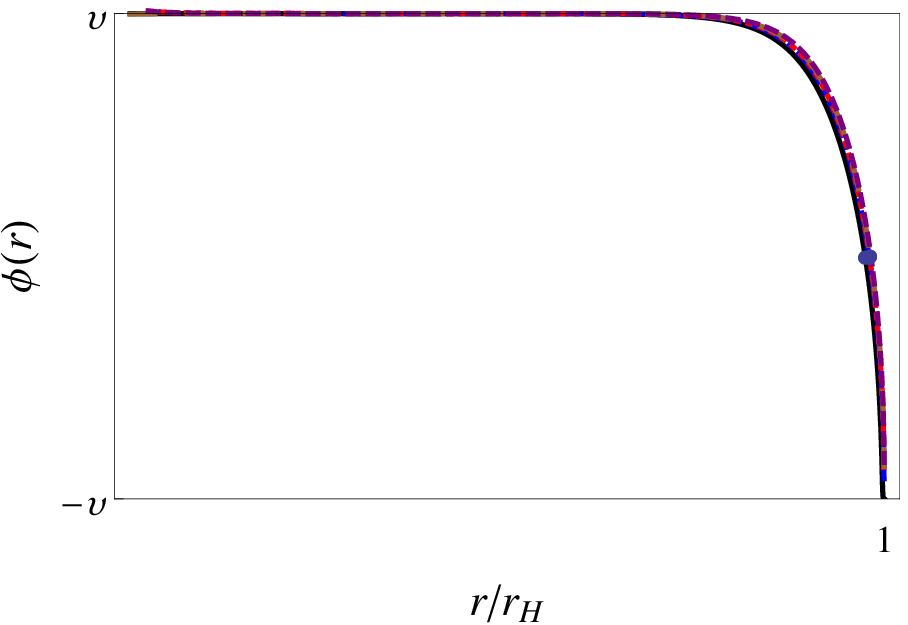}
\\   (c) \hskip 3.0in (d) \\
\includegraphics[height=2.0in]{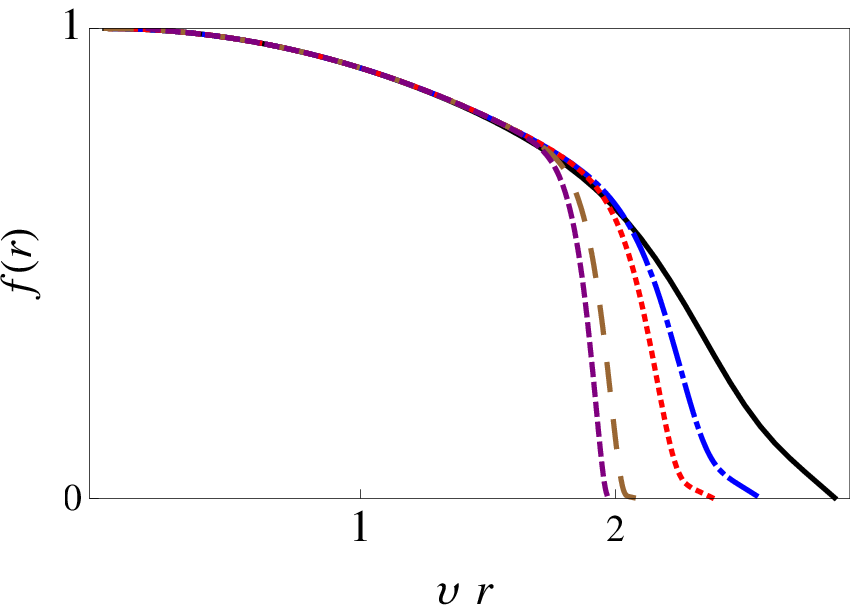}
\hskip .5in
\includegraphics[height=2.0in]{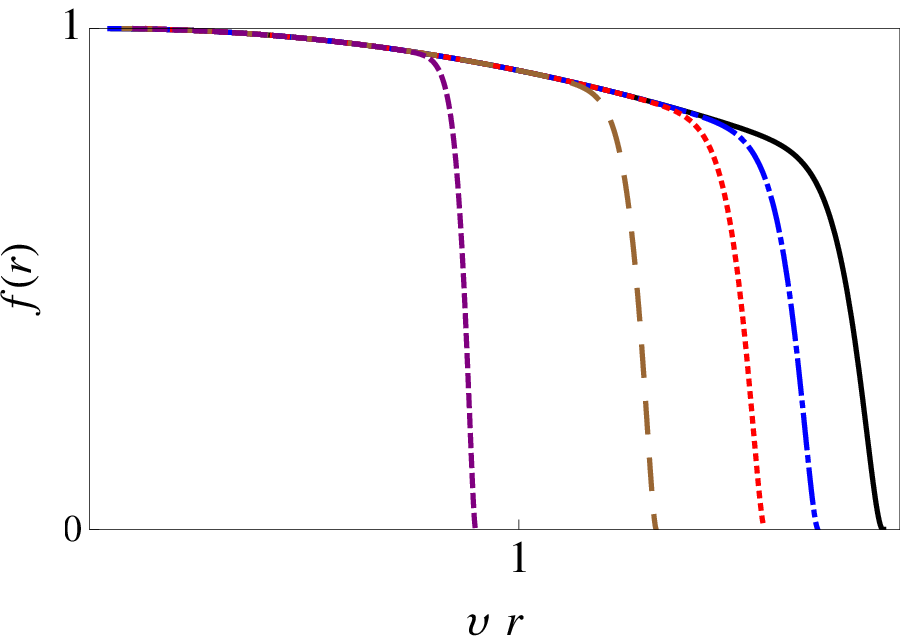}
\\   (e) \hskip 3.0in (f) \\
\end{tabular}
\caption{Variation of the bounce with barrier height.  The panels on
  the left correspond to smaller values of $\alpha$, with the
  short-dashed purple, long-dashed brown, dotted red, dot-dashed blue,
  and solid black lines corresponding to $\alpha$ equal to 1, 3, 5, 12, and 15.
  The right-hand panels correspond to larger values
  of $\alpha$, with the short-dashed purple, long-dashed brown, dotted
  red, dot-dashed blue, and solid black lines corresponding to
  $\alpha$ equal to 15, 20, 25, 40, and 100.  For all of these $U_0=2$,
  $\epsilon=0.3$, and $8\pi G v^2 = 0.126$.  As in the previous
  figures, the black dot indicates the value of the field at the top
  of the barrier.}
\label{varyAlpha}
\end{figure}

Perhaps the most interesting behavior is that which occurs when we
vary $\alpha$, the parameter characterizing the barrier height in the
potential.  Increasing the barrier height increases the surface
tension in the bubble wall.  In flat spacetime this, in turn, requires
that the bubble nucleate with a larger radius.  A second effect of the
increased barrier height is that the bubble wall gets thinner.

Figure~\ref{varyAlpha} shows the effects of varying $\alpha$ on the
O(3)$\times$O(2)-symmetric bounces.  We see two distinct types of
behavior.  For smaller values of $\alpha$, illustrated in the panels
on the left, the scalar field starts close to the true vacuum value at
$r=0$ and then comes exponentially close to the false vacuum by the
time that the horizon is reached.  In contrast with flat spacetime,
the radius at nucleation decreases as $\alpha$ increases.  This is
largely due to the fact that the increased tension in the bubble wall
causes the metric coefficient $f$ to turn over sooner, leading to a
smaller horizon radius.  However, when the field profile is viewed as
a function of $r/r_H$, the result is closer to what the flat spacetime
experience would have led us to expect.  Whether viewed as a function
of $r$ or of $r/r_H$, the bubble wall becomes thinner and steeper with
increasing $\alpha$.

Viewing these results for small $\alpha$, one might be led to expect
that further increases in $\alpha$ would lead to an even steeper
profile that pushed ever closer to the horizon, perhaps becoming a
step function, with an infinitely thin bubble wall, for sufficiently
large $\alpha$.  This is not what happens.  Instead, the
value of the field at the horizon, $\phi(r_H)$, falls visibly short of
the false vacuum by an amount that increases with $\alpha$.  [For the
highest value of $\alpha$ shown here, $\phi(r_H) = -.85 v$.]
Furthermore, the wall does not become thinner.  Indeed, when plotted
as functions of $r/r_H$, the various large-$\alpha$ field profiles are
almost indistinguishable.

\section{Discussion and concluding remarks}

We have studied a class of bounce solutions with O(3)$\times$O(2)
symmetry.  These are curved spacetime analogues of the flat spacetime
bounces describing thermal nucleation of critical bubbles. Following
the interpretation of Ref.~\cite{Brown:2007sd}, the critical bubble
configuration is obtained by taking a constant $\tau$ slice through
the bounce.  This slice is topologically a three-ball, and gives the 
scalar field and metric on a horizon volume.  No reference is made
to quantities beyond the horizon, and the end result of the process is
the creation of one bubble, not two.

Using a general quartic potential as an example, we have explored the
behavior of the bounces in various regions of parameter space.  Either
increasing Newton's constant or raising $V(\phi)$ by a uniform
constant, with the shape of the potential otherwise held fixed, drives
the bounce toward, and eventually merges it with, the Hawking-Moss
solution.  Decreasing the energy difference between the new true
vacuum and the background false vacuum tends to increase the bubble
radius at nucleation, but the effect is hardly as dramatic as in flat
spacetime, even in the case where $\epsilon$ changes sign so that the
bounce is actually describing the nucleation of a false vacuum bubble
in a region of true vacuum.  

The effect of increasing the surface tension in the bubble wall by
raising the potential barrier is particularly notable.  Initially,
this makes the bubble wall thinner and moves it closer to the horizon.  This 
fits one's expectations from the behavior of the thin-wall analysis of
Garriga and Megevand~\cite{Garriga:2003gv,Garriga:2004nm}.  However,
with further increases the two analyses diverge.  In the
thin-wall/membrane case the bubble wall reaches the horizon at a
critical surface tension $\sigma_N$.  By contrast, in our field theory
case the field profile at the bubble wall asymptotes toward one with
nonzero thickness, and in fact shows little variation with increased
surface tension.

To be relevant for tunneling, it is not sufficient that the bounce be
a solution of the Euclidean field equations.  In addition, the
spectrum of fluctuations about the bounce must include one negative
mode, to give an imaginary part to the energy of the false vacuum, but
no more, since additional negative modes would signal that the bounce
corresponded to a path that was not a local minimum of the tunneling
exponent.  There are potentially three types of negative modes for the
O(3)$\times$O(2) symmetric solutions.  Two of these are closely
analogous to the ones associated with thermal bounces in flat
spacetime.  The first is a mode that is independent of the Euclidean
time and corresponds to the radial expansion or contraction of the
critical bubble profile.  We expect that this negative mode will
always be present.  The second class of modes are obtained by
modulating the amplitude of the previous mode with a sinusoidal variation in
the imaginary time.  By analogy with the discussion in
Sec.~\ref{finiteT}, we expect these additional modes to be present if
the ratio of the bubble radius to the inverse temperature is too low.
In order to avoid these, the bubble radius should be comparable to
$r_H$.

The third type of negative mode is closely tied to the four-sphere
topology of the bounce.  It is perhaps easiest to visualize in the
limit where the bubble radius is much less than the horizon radius.
In this limit the shaded region in Fig.~\ref{o3o2-bounce} becomes a
narrow band stretched along the circumference of the bounce.  Moving
this true vacuum region to either side reduces its length in the
$\tau$ direction and therefore reduces its action.  This mode may be
avoided if the bubble radius is much larger and the bubble wall is no
longer thin but instead has a tail extending out toward the horizon.
This is the same region of parameter space as is required to avoid
additional negative modes of the second type.

\begin{figure}
\centering
\begin{tabular}{cc}
\includegraphics[height=1.95in]{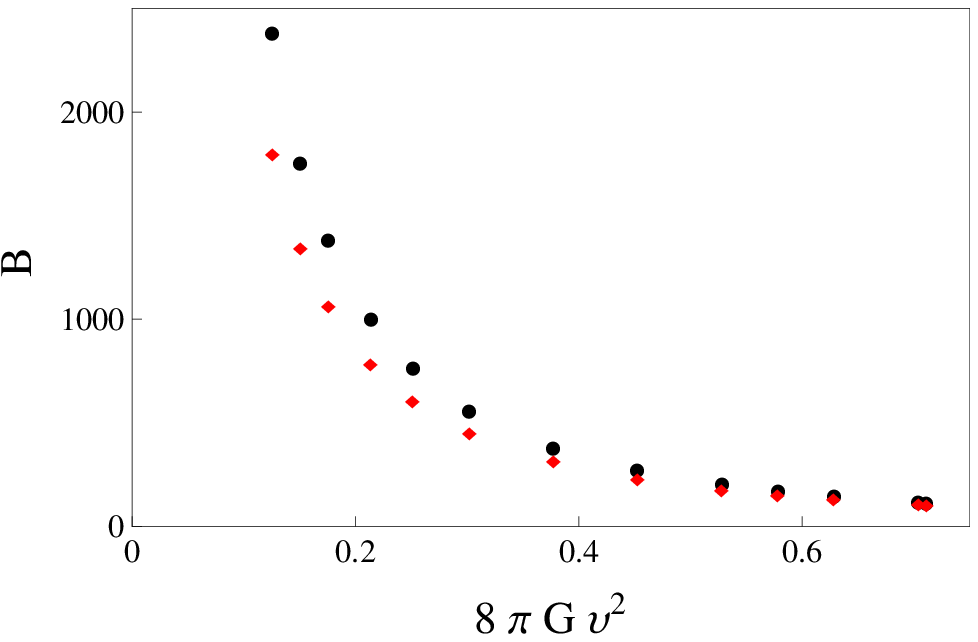}
\hskip .18in
\includegraphics[height=1.95in]{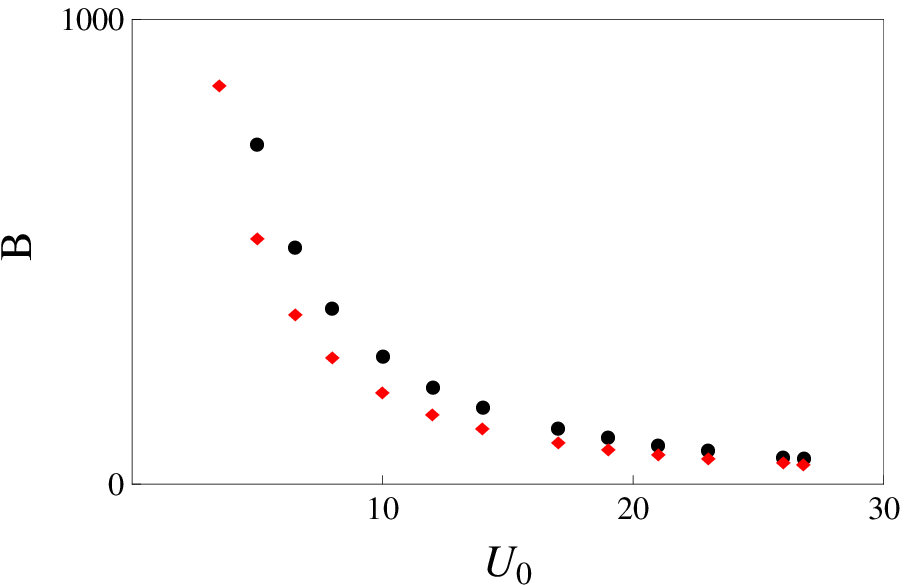}
\\   (a) \hskip 3.0in (b) \\
\includegraphics[height=1.95in]{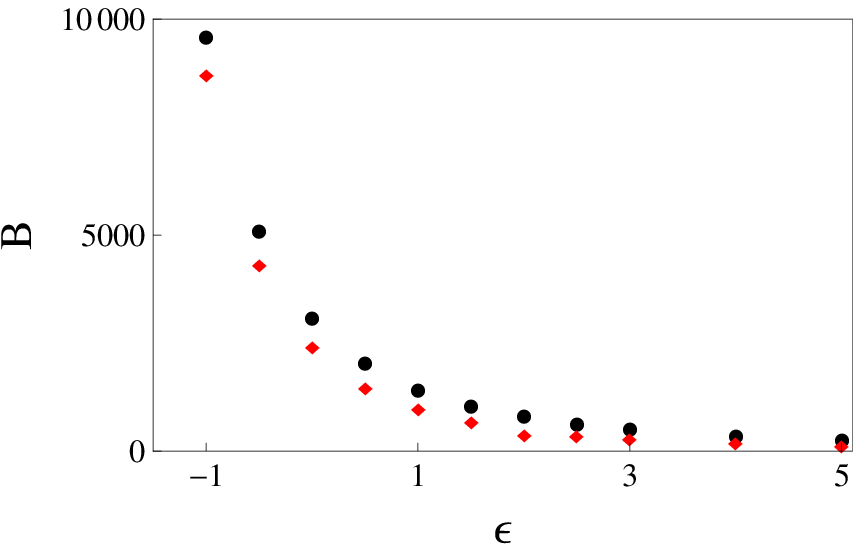}
\hskip .18in
\includegraphics[height=1.95in]{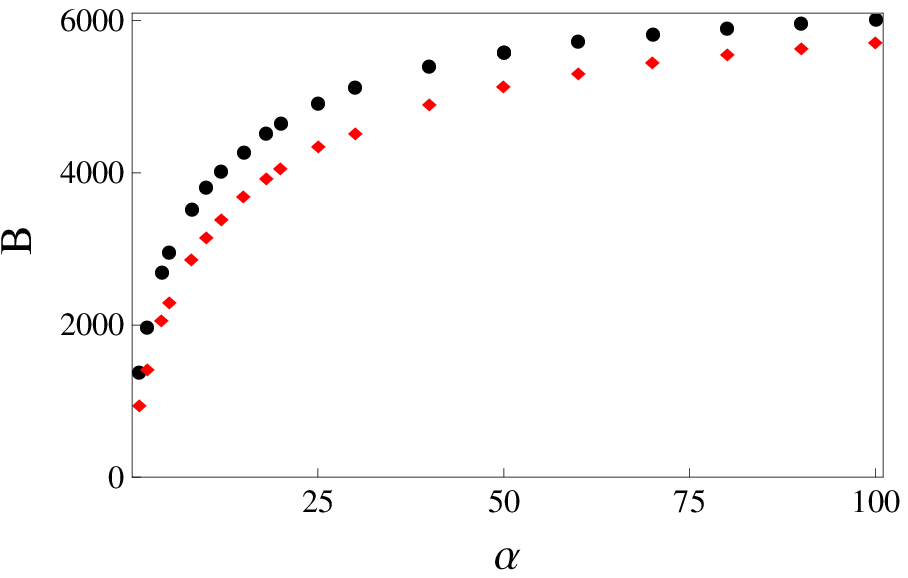}
\\   (c) \hskip 3.0in (d) 
\end{tabular}
\caption{Comparison of the tunneling exponents for the CDL bounce (red
  diamonds) and the O(3)$\times$O(2)-symmetric bounce (black circles)
  as one parameter is varied with the other three held fixed.  When
  held fixed, $U_0=2$, $\alpha=3$, $\epsilon=0.3$, and $8\pi G
  v^2=0.126$. }
\label{actionPlots}
\end{figure}

Finally, even if the O(3)$\times$O(2)-symmetric bounce has only a
single negative mode, it is of practical relevance only if its action
is less than that of the CDL bounce.  Indeed, one of the goals of our
investigation was to test the conjecture that the bounce with minimum
action has O(4) symmetry.  It was already shown in
Ref.~\cite{Garriga:2004nm} that in the thin-wall approximation the
O(3)$\times$O(2)-symmetric bounce always has a higher action than the
corresponding CDL action.  Our numerical methods allow us explore the
much larger parameter space that is allowed when one goes beyond this
approximation.  The results are summarized in Fig.~\ref{actionPlots}.
Each of the four panels corresponds to a one-dimensional path through
this parameter space.  We see that the tunneling exponent for the
O(3)$\times$O(2) is never less than that of the CDL bounce.  The two
approach and become equal at large values of $G$ and of $U_0$. This
can be understood by recalling that the O(3)$\times$O(2)-symmetric
bounce reduces to the Hawking-Moss solution when $V''/H^2 \le 10$ at
the top of the barrier [see Eq.~(\ref{HMcondition})].  For $4<V''/H^2
\le 10$ there is always a CDL bounce whose action is less than that of
the Hawking-Moss solution, but at $V''/H^2 =4$ the CDL solution merges
into the Hawking-Moss.

These numerical tests are, of course, not conclusive.  Although we
have no reason to expect it, there could be unexplored regions of our
parameter space where the action is less than the CDL action.  More
generally, there may be other choices of the potential for which the
CDL bounce is not optimal.  Conclusive analytical results on this
issue would be welcome.

\begin{acknowledgments}

We thank Alex Vilenkin for bringing the work of Garriga and Megevand
to our attention, and are grateful to Adam Brown and Alex Dahlen for
enlightening conversations.  This work was supported in part by the
U.S.~Department of Energy.  E.J.W. thanks the Aspen Center for
Physics, supported by NSF Grant \#1066293, for hospitality during the
completion of this work.

\end{acknowledgments}

\end{document}